\documentclass[aps,prl,twocolumn,showpacs,superscriptaddress]{revtex4}
\usepackage{graphicx}

\begin{document}

\title{Reversal of Thermal Rectification in Quantum Systems}
     \author{Lifa~Zhang}
     \affiliation{Department of Physics and Centre for Computational Science and Engineering,
     National University of Singapore, Singapore 117546, Republic of Singapore }
     \author{Yonghong~Yan}
     \affiliation{D\'{e}partement de Chimie, B6c, Universit\'{e} de Li\`{e}ge, B4000 Li\`{e}ge, Belgium}
     \author{Chang-Qin Wu}
     \affiliation{Department of Physics, Fudan University, Shanghai 200433, People's Republic of China}
     \affiliation{Department of Physics and Centre for Computational Science and Engineering,
     National University of Singapore, Singapore 117546, Republic of Singapore }
     \author{Jian-Sheng~Wang}
     \affiliation{Department of Physics and Centre for Computational Science and Engineering,
     National University of Singapore, Singapore 117546, Republic of Singapore }
     \author{Baowen~Li}
      \altaffiliation{Electronic address: phylibw@nus.edu.sg}
       \affiliation{Department of Physics and Centre for Computational Science and Engineering,
     National University of Singapore, Singapore 117546, Republic of Singapore }
     \affiliation{NUS Graduate School for Integrative Sciences and Engineering,
      Singapore 117456, Republic of Singapore}
\date{11 Nov 2009}

\begin{abstract}
{We study thermal transport in anisotropic Heisenberg spin chains
using the quantum master equation. It is found that thermal
rectification changes sign when the external homogeneous magnetic
field is varied. This reversal also occurs when the magnetic field
becomes inhomogeneous. Moreover, we can tune the reversal of rectification by  temperatures of the heat baths, the anisotropy and size of the spin chains. }
\end{abstract}

\maketitle

Considerable progress has been made both theoretically and
experimentally on thermal transport in micro and nano scale in last
decade\cite{Dhar}. It has been found that similar to electrons, the
heat due to phonons can be used to carry and process information
\cite{Wang2008}.  In particular, several conceptual thermal devices
have been proposed, such as thermal rectifiers \cite{rectifiers},
thermal transistors \cite{transistor}, thermal logical gates
\cite{logicgate}, thermal memory\cite{memory}, some molecular level
thermal machines \cite{segal2006,marathe2007}, and thermal ratchet
\cite{LiNB}. Much work has also been done in quantum heat transport
of nanostructures \cite{wangjs2008}, and spin systems
\cite{castella1995,saito1996,zotos1997,naef1998,louis2003,
heidrich2005,kordonis2006,hess2007,sologubenko2007,sologubenko2008,zhang2008,yan2008}
where the magnetic field is another degree of freedom to control
heat flow. Indeed, it is demonstrated that thermal rectification and
negative differential thermal resistance are observable in quantum
spin chain by applying a nonuniform magnetic field \cite{yan2009}.

In this paper, we would like to concentrate on the thermal
rectification in a quantum spin model. Our primary interest is to
understand how far we can control the heat flow by tuning the system
parameters such as magnetic field, system size, configurations of
the system etc. In particular, we would like to see whether the reversal
of thermal rectification, which has been observed in classical
system\cite{hu2006}, can happen in such a quantum system.

We consider a Heisenberg spin-1/2 chain,  whose Hamiltonian reads
\begin{equation}
 H =  - \sum\limits_{i = 1}^{N - 1} {(J_x \sigma _i^x \sigma _{i + 1}^x  + J_y \sigma _i^y \sigma _{i + 1}^y  + J_z \sigma _i^z \sigma _{i + 1}^z ) - \sum\limits_{i = 1}^N {h_i \sigma _i^z } },
\end{equation}
where $N$ is the number of spins, the operators $\sigma
_i^x,\,\sigma _i^y$ and $\sigma _i^z$ are the Pauli matrices for
the i\textit{th} spin, $J_x, \, J_y$ and $J_z$ are the coupling
constants between the nearest-neighbor spins, and $h_i$ is the
magnetic field strength at the i\textit{th} site. We set
$J_x=J(1+\gamma), \, J_y=J(1-\gamma),\, J_z=J$ to consider the
anisotropy in $x-y$ plane, where $\gamma$ is the anisotropy
parameter, and $J=1$, without loss of generality. Figure 1
\begin{figure}
\includegraphics[width=2.5 in,angle=0]{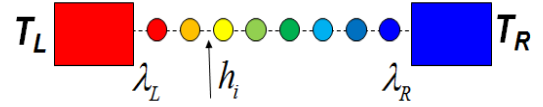}%
\caption{ \label{fig1model}(color online) A schematic
representation of the model with size $N=8$. The spin chain is
connected to two phonon baths with coupling $\lambda_L$ and
$\lambda_R$. The phonon baths are at different temperature $T_L $
and $ T_R$. A magnetic field is applied to the spin chain. }
\end{figure}
shows a schematic representation of this model.

The total Hamiltonian including two baths is
\begin{equation}
 H_{tot}  = H + H_B  + H_I.
\end{equation}
Here $H_B$ is the Hamiltonian of the heat baths
$ H_B  = \sum\limits_{K = L,R} {H_B^K } $, $H_B^K  = \sum\limits_{j \in K} {\omega _j b_j^ +  b_j } $, where $b_j^+$ and $b_j$ are phonon creation and annihilation operators  with the  mode $\omega _j$;
 $H_I$ is the interaction between the spin chain and phonon heat baths, $
 H_I  = \sum\limits_{K = L,R} {X_K  \otimes Y_K} $, where $
 X_L  = \sigma _1^x ,X_R  = \sigma _N^x $, and $Y_K$ is bath operator $ Y_K = \sum\limits_{j \in K} (c_j b_j^+   + c_j^* b_j) $.

\begin{figure}[h]
\includegraphics[width=3.2 in,angle=0]{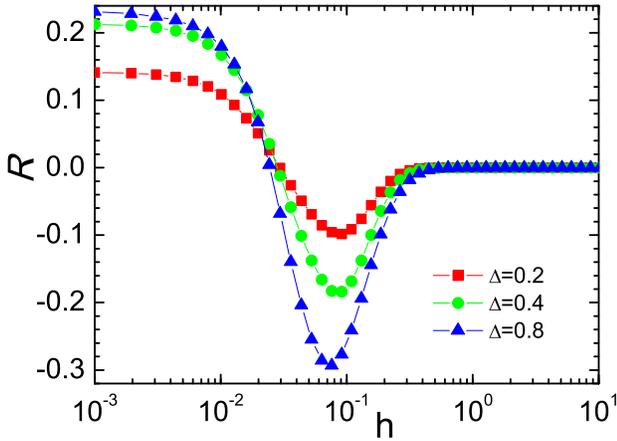}%
\caption{ \label{fig2recthz}  (color online)  Rectification as a
function of the  magnetic field in a chain of size $N=6$. The
temperature of the heat bath are $T_+=T_0(1+\Delta)$ and
$T_-=T_0(1-\Delta)$, where $T_0=0.1$ is the mean temperature. The
coupling between the spin chain and heat bath are $\lambda_L=0.16$
and $\lambda_R=0.04$, anisotropy parameter $\gamma=0.1$. The square,
circle and triangular correspond to $\Delta=0.2$, 0.4 and 0.8,
respectively.}
\end{figure}
We use the quantum master equation (QME) method
(Refs.\cite{yan2009,kubo1991,saito2003,segal2005,michel2008}) to
study heat conduction in this model. By tracing out the baths within
the Born-Markovian approximation, we obtain the equation of motion
for the reduced density matrix of the system ($\hbar=1$),
\begin{equation}
\frac{d}{{dt}}\rho  =  - i[H,\rho ] + \mathcal{L}_L\rho  + \mathcal{L}_R \rho,
\end{equation}
where $\mathcal{L}_L\rho$ and $\mathcal{L}_R\rho$ are  the
dissipative terms due to the coupling with the left and right heat
bath. $\mathcal{L}_L \rho $ is given by $\mathcal{L}_L \rho  = [
\mathcal{X}_L \rho ,X_L ] + {\rm h.c.}$ and  $\mathcal{L}_R\rho$ can
be given in the similar way. Here the operator $\mathcal{X}_L$ can
be written as
\begin{equation}
\langle m|\mathcal{X}_L|n\rangle=\lambda_L \varepsilon_{m,n} N_L(\varepsilon_{m,n})\langle m|X_L|n\rangle,
\end{equation}
where $\varepsilon_{m,n} \equiv \varepsilon_m-\varepsilon_n$ and
$N_L(\varepsilon_{m,n})=(e^{\varepsilon_{m,n}/T_L}-1)^{-1}$ is the
Bose distribution ($k_B=1$) with $T_L$ being the temperature of left
heat bath. $|n\rangle$ and $\varepsilon_n$ are the eigenstates and
eigenvalues  of the spin chain system. The bath spectrum function we
used is of an Ohmic type. Assuming that the temperature is high
enough to make dephasing fast \cite{segal2005},  we can solve the
resulting kinetic equations of the state probabilities  numerically.
The evolution time is chosen long enough such that the final density
matrix reaches a steady state $\rho_{\rm st}$, that is,
$\dot{\rho}_{nn}=0$. In the steady state,
$\sum_{n}\varepsilon_n\dot{\rho}_{nn} =J_L+J_R=0$, then we can get
the heat current  as $J=J_L=-J_R, \;{\rm if}\; T_L>T_R$. To quantify
the rectification efficiency, we define rectification $R$ as: $R =
(J_+ - J_-)/ {\rm max }\{J_+, J_-\}$, where  the forward heat flux
$J_+$  is the heat current (from left to right) when the bath at
higher temperature $T_+$  is connected to the left end of the chain
and  the backward flux $J_-$ is the heat current (from right to
left) when the left end of the chain is in contact with the bath at
lower temperature $T_-$.
\begin{figure}[h]
\includegraphics[width=3.5 in,angle=0]{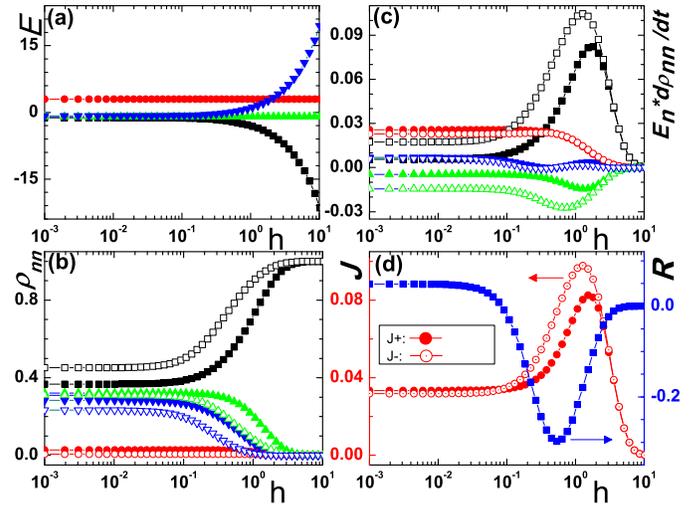}%
 \caption{\label{fig3mech2spin} (color online) Analysis of the reversal of rectification in a  two-spin system. Parameters: $ T_0=1.0,\,\Delta=0.8$, $\lambda_L=0.16$, $\lambda_R=0.04$, $\gamma=0.1$.
(a) The eigenvalues $E_n$ vs magnetic field $h$. (b) The density
$\rho_{nn}$ (probability of each eigenstate) vs $h$. (c) The product
of $E_n$ and the derivative of $\rho_{nn}$ from high temperature
heat bath, which is the contribution to heat flux from each
eigenstate. In (a), (b), and (c), the square, circle, up triangle,
and down triangle correspond to eigenstate 1, 2, 3, and 4,
respectively. In (b) and (c), the solid and hollow symbols
correspond to forward and backward thermal transport, respectively.
(d) The heat fluxes $J_+,\,J_-$ vs magnetic field $h$ (left scale).
The square curve shows rectification effect $R$ vs magnetic
field $h$ (right scale). }
\end{figure}

The quantum spin chain is a nonlinear system; if we introduce
asymmetry to the system, then it may show the rectification effect.
We connect the spin chain to the two heat baths by different
couplings $\lambda_L=0.16$ and $\lambda_R=0.04$. As is shown in
Fig.~\ref{fig2recthz}, when the applied magnetic field increases,
the rectification changes sign. This phenomenon is not observed in
electronic counterpart. From Fig.~\ref{fig2recthz}, we can see that
rectification $R$ can be positive, zero, or negative, depending on
the magnetic field $h$. The behavior of $R$ remains similar for
different $\Delta$, the temperature difference of the baths.

The mechanism of thermal rectification reversal can be understood
from a two-spin system which has four eigenstates, whose
contribution to heat transport can be seen clearly in
Fig.~\ref{fig3mech2spin}. Fig.~\ref{fig3mech2spin}(a) shows the
eigenvalues $E_n$ vs magnetic field $h$, where two eigenvalues do
not change with the magnetic field. Fig.~\ref{fig3mech2spin}(b)
shows the steady density $\rho_{nn}$ as a function of magnetic
field. From this figure, we find that when the magnetic field is
weak, the ground state and the excited states have some probability
to be occupied  (the ground state has the largest probability). If
the magnetic field increases, the probability of the ground state
becomes larger; after a certain value, the probability of ground
state is close to one, while others are zero. The heat current from
the contribution of each eigenstate  is shown in
Fig.~\ref{fig3mech2spin}(c). When the magnetic field is weak, each
state  contributes to the heat flux either  positively or
negatively. We find that the contribution from ground state in the
backward thermal transport is larger than the forward case. But the
contributions from other excited states have negative effect;
therefore the total effect is the forward heat flux larger than the
backward flux, that is, the rectification is positive. When the
magnetic field increases over a certain value,  the ground state
will dominate the heat transport, and the contribution from other
states decrease. At the time the total heat flux will have the
similar behavior with the contribution of ground state: the backward
flux is larger than the forward one; both of them increase first and
decrease to zero at last, when the system will stay in the ground
state. Therefore, the rectification changes from positive to
negative, then from negative to zero at last, which can be seen in
Fig.~\ref{fig3mech2spin}(d). In short, the heat flux from the
contribution of ground state is larger when the more weakly coupled
reservoir is hotter, that is, the backward one is larger than the
forward one; however for the flux from excited states, the forward
one is larger than the backward one. The rectification is determined
by the competition of the contributions to heat flux from the ground
state and the excited states.

\begin{figure}
\includegraphics[width=3.2 in,angle=0]{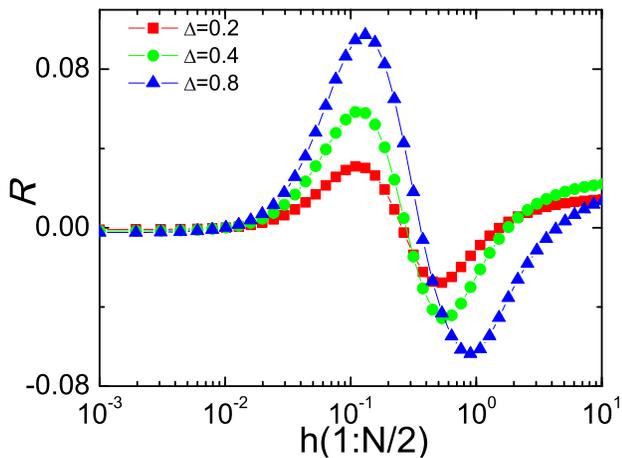}%
 \caption{\label{fig4rectinhz} (color online) The rectification $R$ vs nonuniform magnetic field. The chain size is  $N=6$, $h(1:N/2)$ is the magnetic field applied to site 1 to $N/2$, and others are zero, that is, $h(N/2+1:N)=0$. Here $\gamma=0$, $T_0=0.1$ and $\lambda_L=\lambda_R=0.1$. The square, circle and triangular correspond to $\Delta=0.2$, 0.4 and 0.8 respectively.}
\end{figure}

Applying a nonuniform magnetic field to the spin chain is another
possible  way to introduce  asymmetry to the system, and the
system can also exhibit rectification. Fig.~\ref{fig4rectinhz}
demonstrates this phenomenon in such a system. The rectification
is zero when the magnetic field is zero because of no asymmetry in
the system. When a weak magnetic field is applied to the left half
part of the spin chain, the rectification becomes positive. When
magnetic field increases, the energy difference increases for the
left part, which enlarges the rectification. If the field
increases further, then the rectification reverses.  For different
temperature difference, the rectification has similar effect but
different magnitude.

\begin{figure}[h]
\includegraphics[width=3.2 in,angle=0]{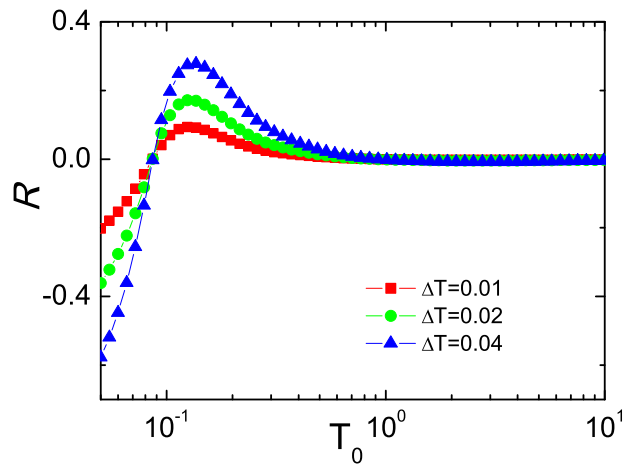}%
\caption{\label{fig5rectT}  (color online) Rectification as a
function of  mean temperature $T_0$. The size of spin chain is
$N=6$. The temperature of the heat bath are $T_+=T_0+\Delta T$ and
$T_-=T_0-\Delta T$. Here, $\lambda_L=0.16$, $\lambda_R=0.04$,
$\gamma=0.1$ and $h=0.01$. The square, circle and triangular
correspond to $\Delta T=0.01$, $0.02$ and $0.04$
respectively.}
\end{figure}

\begin{figure}[h]
\includegraphics[width=3.2 in,angle=0]{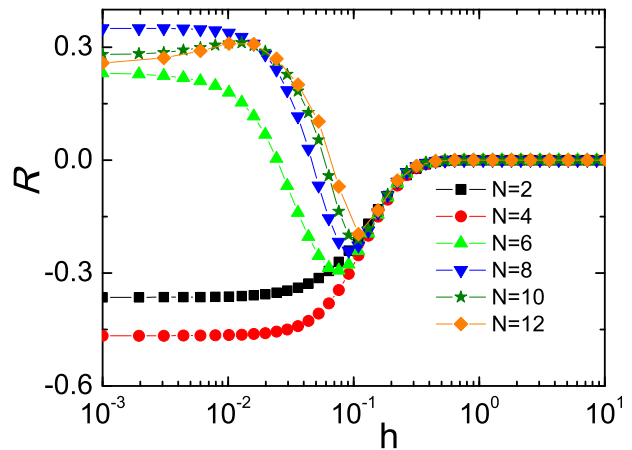}%
 \caption{\label{fig6rectsize} (color online) The rectification
changes with  magnetic field for the spin chain with different
sizes. The curves of square, circle and  up triangle, and down
triangle, star and diamond correspond to $N=2$, 4, 6, 8, 10 and 12 respectively. Here the parameters are: $T_+=0.18$ and $T_-=0.02$, $\lambda_L=0.16$,
$\lambda_R=0.04$ and $\gamma=0.1$.}
\end{figure}

From the above discussions, we find that the rectification can
change sign with the applied magnetic field. Indeed, an increase of
mean temperature  can also induce  reversal of rectification.
Fig.~\ref{fig5rectT} shows that the rectification as a function of
mean temperature $T_0$. Here we keep the temperature difference
fixed, and increase the mean temperature of the heat baths; the
rectification changes sign from negative to positive. When the
temperature is very low, the ground state dominates the thermal
transport, the backward flux is larger than the forward one; if we
raise the temperature, more excited states will contribute to the
heat flux and gradually control the thermal transport, when the
rectification changes to positive.  For different temperature
difference, the rectification changes sign at almost the same mean
temperature.
\begin{figure}[h]
\includegraphics[width=3.2 in,angle=0]{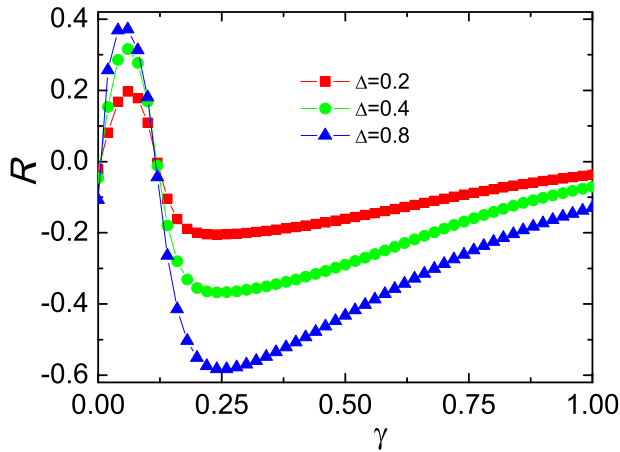}%
 \caption{\label{fig7rectgama} (color online)
The rectification changes with  anisotropy in the spin chain. The
chain size is  $N=6$. The square, circle and
triangular correspond to $\Delta=0.2$, 0.4 and 0.8 respectively.
Here the parameters are: $T_+=0.18$ and $T_-=0.02$,
$\lambda_L=0.16$, $\lambda_R=0.04$ and $h=0.01$.}
\end{figure}

Rectification can change sign with the external parameter, such as
magnetic field and the temperatures of the heat baths. In our study,
we find that thermal rectification can reverse with the properties
of the spin chain itself, such as the size and the anisotropy of the
spin chain. Fig.~\ref{fig6rectsize} shows the rectification changes
with the magnetic field for different size cases. The rectification
effect behaves differently for different size. In
Fig.~\ref{fig6rectsize}, there is no reversal of rectification for
small size cases $N=2$ and $N=4$; but for larger size cases
$N=6, 8, 10$ and $12$, it shows reversal of rectification.
Fig.~\ref{fig7rectgama} shows the rectification reverses with the
anisotropy of the spin chain. In the weak anisotropy range, the
rectification coefficient is positive, but it changes to negative
when the anisotropy is strong. During the changing of anisotropy,
the forward total flux is larger than the backward one at first, and
then reverses, although the heat contribution from ground state in
the backward transport is always larger than that in the forward
one.

In conclusion, we have studied thermal rectification in quantum
spin-chain systems by using quantum master equations. It has been
shown that rectification can change sign when the magnetic field,
temperature, the anisotropy, and the system size change.  Although
the reversal of rectification is complicated parameter-dependent, it
is believed to be a universal phenomenon for the thermal transport
in one-dimension systems.

We thank Jiang Jinwu and Ren Jie for fruitful discussions. L.Z. and B.L. are supported by the grant R-144-000-203-112 from
Ministry of Education of Republic of Singapore. J.-S.W.
acknowledges support from a faculty research grant
R-144-000-173-112/101 of NUS. C.Q.W. is supported by the NSF of
China and the MOE of China (through Project B06011).

\end{document}